\pdfoutput=1

\documentclass[twocolumn]{svjour3}

\smartqed

\usepackage{graphicx}
\usepackage{amssymb}
\usepackage{epsf}
\usepackage{txfonts}
\usepackage{natbib}
\usepackage[unicode]{hyperref}

\hypersetup{pdftitle = The title of my PDF, pdfauthor = My name, pdfsubject= The subject, pdfkeywords = keyword1 keyword2 keyword3}
\hypersetup{colorlinks = true, linkcolor = blue, anchorcolor = red, citecolor = blue, filecolor = red, pagecolor = red, urlcolor = blue}

\journalname{Nonlinear Dynamics}

\begin{document}

\title{Classifying orbits in the restricted three-body problem}

\author{Euaggelos E. Zotos}

\institute{Department of Physics, School of Science, \\
Aristotle University of Thessaloniki, \\
GR-541 24, Thessaloniki, Greece \\
Corresponding author's email: {evzotos@physics.auth.gr}
}

\date{Received: 4 March 2015 / Accepted: 15 June 2015 / Published online: 25 June 2015}

\titlerunning{Classifying orbits in the restricted three-body problem}

\authorrunning{Euaggelos E. Zotos}

\maketitle

\begin{abstract}

The case of the planar circular restricted three-body problem is used as a test field in order to determine the character of the orbits of a small body which moves under the gravitational influence of the two heavy primary bodies. We conduct a thorough numerical analysis on the phase space mixing by classifying initial conditions of orbits and distinguishing between three types of motion: (i) bounded, (ii) escape and (iii) collisional. The presented outcomes reveal the high complexity of this dynamical system. Furthermore, our numerical analysis shows a remarkable presence of fractal basin boundaries along all the escape regimes. Interpreting the collisional motion as leaking in the phase space we related our results to both chaotic scattering and the theory of leaking Hamiltonian systems. We also determined the escape and collisional basins and computed the corresponding escape/collisional times. We hope our contribution to be useful for a further understanding of the escape and collisional mechanism of orbits in the restricted three-body problem.


\end{abstract}

\section{Introduction}
\label{intro}

The planar circular restricted three-body problem has played in the past a very essential role in many different fields of dynamical astronomy and celestial mechanics. For example, the modern applications to space mechanics and dynamics are probably even more cogent than the classical applications. Today numerous aspects in space dynamics are of paramount importance and of great interest. The applications of the restricted three-body problem create the basis of most of the lunar and planetary theories used for launching artificial satellites in the Earth-Moon system and in solar system in general.

Over the last decades a huge amount of research work has been devoted on the subject of escaping particles from open dynamical systems. Especially the issue of escape in Hamiltonian systems is a classical problem in nonlinear dynamics (e.g., [\citealp{C90}, \citealp{CK92}, \citealp{CKK93}, \citealp{CHLG12}, \citealp{STN02}]). It is well known, that several types of Hamiltonian systems have a finite energy of escape. For values of energy lower than the escape energy the equipotential surfaces of the systems are closed which means that orbits are bound and therefore escape is impossible. For energy levels above the escape energy on the other hand, the equipotential surfaces open and exit channels emerge through which the particles can escape to infinity. The literature is replete with studies of such ``open" Hamiltonian systems (e.g., [\citealp{BBS09a}, \citealp{BBS09b}, \citealp{EP14}, \citealp{KSCD99}, \citealp{NH01}, \citealp{STN02}, \citealp{Z14a}, \citealp{Z14ip}]). At this point we should emphasize, that all the above-mentioned references on escapes in Hamiltonian system are exemplary rather than exhaustive, taking into account that a vast quantity of related literature exists.

Nevertheless, the issue of escaping orbits in Hamiltonian systems is by far less explored than the closely related problem of chaotic scattering. In this situation, a test particle coming from infinity approaches and then scatters off a complex potential. This phenomenon is well investigated as well interpreted from the viewpoint of chaos theory (e.g., [\citealp{BGOB88}, \citealp{E87}, \citealp{J87}, \citealp{JLS99}, \citealp{JMS95}, \citealp{JP89}, \citealp{JR90}, \citealp{JS87}, \citealp{JT91}, \citealp{SASL06}, \citealp{SSL07}, \citealp{SS08}, \citealp{SHSL09,SS10}]). Chaotic scattering has also been applied in the astrophysical context of many aspects such as scattering off black holes (e.g., [ \citealp{A97}, \citealp{dML99}]) and three-body stellar systems (e.g., [\citealp{BTS96}, \citealp{BST98}, \citealp{H83}]). The related invariant manifolds of the chaotic saddle are directly associated with the chaotic dynamical behavior (e.g., [\citealp{O02}]). In particular, the chaotic saddle is defined as the intersection of its stable and unstable manifolds [\citealp{S14}], while hyperbolic and non-hyperbolic chaotic saddles may occur in dynamical systems (e.g., [\citealp{LGB93}]). More details on the issue of chaotic scattering and escape from chaotic systems can be found to the recent reviews [\citealp{SS13}] and [\citealp{APT13}], respectively.

In open Hamiltonian systems an issue of paramount importance is the determination of the basins of escape, similar to basins of attraction in dissipative systems or even the Newton-Raphson fractal structures. An escape basin is defined as a local set of initial conditions of orbits for which the test particles escape through a certain exit in the equipotential surface for energies above the escape value. Basins of escape have been studied in many earlier papers (e.g., [\citealp{BGOB88}, \citealp{C02}, \citealp{KY91}, \citealp{PCOG96}]). The reader can find more details regarding basins of escape in \citealp{C02}, while the review [\citealp{Z14ip}] provides information about the escape properties of orbits in a multi-channel dynamical system of a two-dimensional perturbed harmonic oscillator. The boundaries of an escape basins may be fractal (e.g., [\citealp{AVS09}, \citealp{BGOB88}]) or even respect the more restrictive Wada property (e.g., [\citealp{AVS01}]), in the case where three or more escape channels coexist in the equipotential surface.

One of the most characteristic Hamiltonian systems of two degrees of freedom with escape channels is undoubtedly the well-known H\'{e}non-Heiles system [\citealp{HH64}]. A huge load of research on the escape properties of this system has been conducted over the years (e.g., [\citealp{AVS01}, \citealp{AJ03}, \citealp{AVS03}, \citealp{BBS08}, \citealp{BBS09a}]). Escaping orbits in the classical Restricted Three-Body Problem (RTBP) is another typical example (e.g., [\citealp{N04}, \citealp{N05}, \citealp{dAT14}, \citealp{Z15}]). Furthermore, escaping and trapped motion of stars in stellar systems are an another issue of great importance. In a recent article [\citealp{Z12}], we explored the nature of orbits of stars in a galactic-type potential, which can be considered to describe local motion in the meridional $(R,z)$ plane near the central parts of an axially symmetric galaxy. It was observed, that apart from the trapped orbits there are two types of escaping orbits, those which escape fast and those which need to spend vast time intervals inside the equipotential surface before they find the exit and eventually escape. Furthermore, the chaotic dynamics within a star cluster embedded in the tidal field of a galaxy was explored in [\citealp{EJS08}]. In particular, by scanning thoroughly the phase space and obtaining the basins of escape with the respective escape times it was revealed, that the higher escape times correspond to initial conditions of orbits near the fractal basin boundaries.

In the present paper we continue the work initiated in [\citealp{N04}] and [\citealp{N05}] following similar numerical techniques. Although the present paper is an extension of the work presented in [\citealp{N05}], it contains several novel results since we consider more cases and we obtain numerical results for a wider set of the involved parameters. In particular, in Section \ref{numres} we investigate four cases regarding the value of the mass parameter $\mu$. In [\citealp{N05}] however, the same four cases were studied just for only one energy level. In our work on the other hand, we explore for every case three different energy levels which correspond to three different Hill's region. Therefore all the plots presented in Section \ref{numres} are completely new and they can not be regarded as reproductions. Moreover, in Section \ref{over} we explore four new cases regarding the mass parameter ($\mu = 1/4$, $\mu = 1/6$, $\mu = 1/7$ and $\mu = 1/8$) which have not explored in [\citealp{N05}]. In addition to the classical $(x, E)$ plane we introduce a new type of plane which is the $(y, E)$ plane. Furthermore, we present the evolution of the percentages of all types of orbits as a function of the total orbital energy. At this point we should emphasize that, as far as we know, this is the first time that a statistical analysis of the evolution of the percentages is examined in the case of the restricted three-body problem. Therefore, taking all the above points into consideration we believe that our paper makes a significant contribution to the field of the restricted three-body problem.

The paper is organized as follows: In Section \ref{mod} we introduce the considered dynamical model and we present its properties along with some necessary details. All the computational methods we used in order to determine the character of orbits are described in Section \ref{cometh}. In the following Section, we conduct a thorough numerical investigation revealing the overall orbital structure (bounded regions and basins of escape/collision) of system and how it is affected by the total orbital energy considering several cases regarding the value of the mass ratio. In Section \ref{over} a general overview is provided showing in more detail the influence of the value of the energy. Our paper ends with Section \ref{disc}, where the discussion and the conclusions of this research are given.

\section{Details of the dynamical model}
\label{mod}

\begin{figure}[!tH]
\centering
\resizebox{\hsize}{!}{\includegraphics{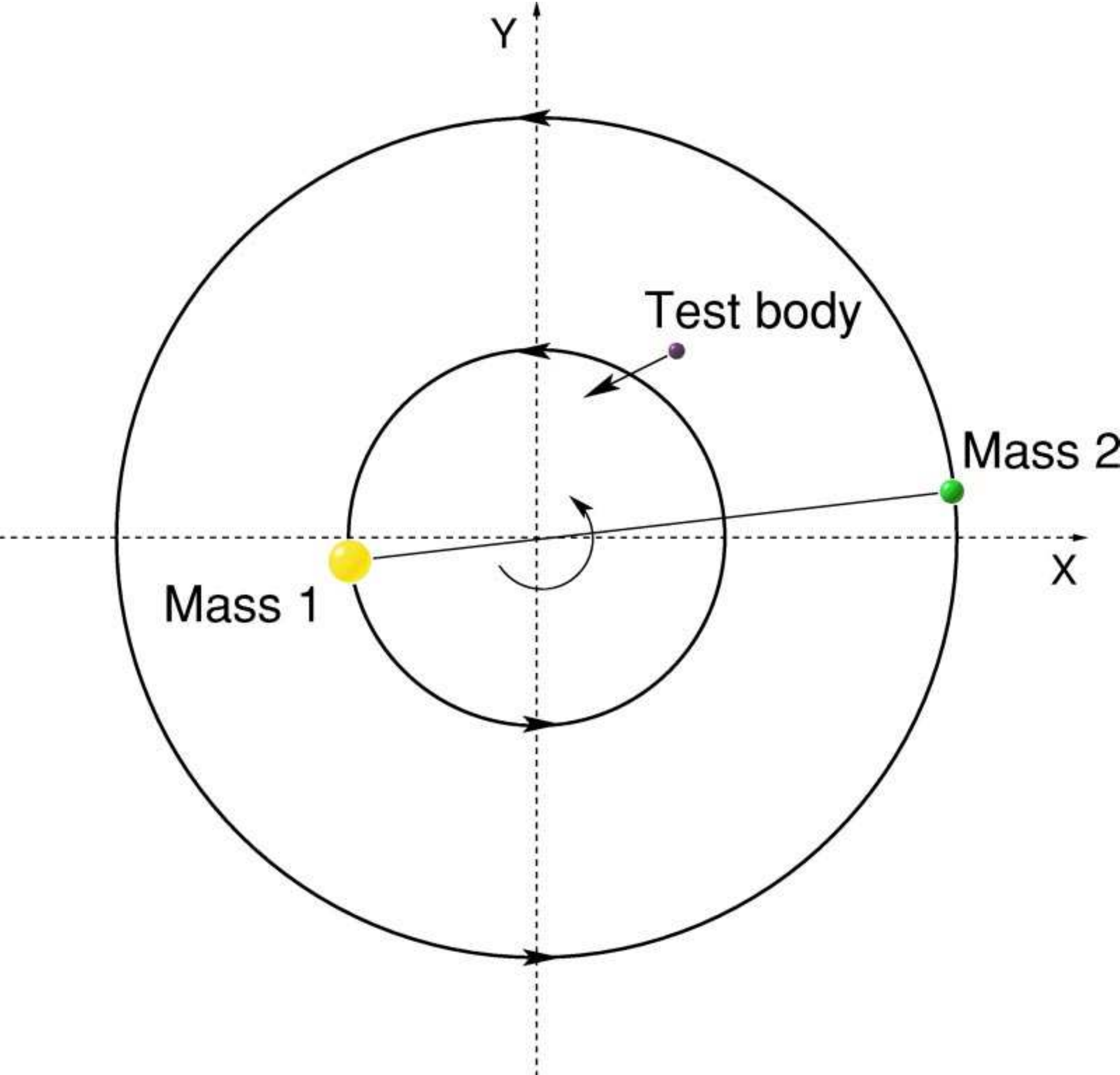}}
\caption{Schematic picture of the planar circular restricted three-body problem.}
\label{rtbp}
\end{figure}

The aim of this work is to investigate the properties of motion in the planar circular restricted three-body problem (PCRTBP). The two primaries move on circular orbits with the same Kepler frequency around their common center of gravity, which is assumed to be fixed at the origin of the coordinates. The third body (test particle with mass much smaller than the masses of the primaries) moves in the same plane under the gravitational field of the two primaries (see Fig. \ref{rtbp}). The non-dimensional masses of the two primaries are $1-\mu$ and $\mu$, where $\mu = m_2/(m_1 + m_2)$ is the mass ratio.

We choose as a reference frame a rotating coordinate system where the origin is at $(0,0)$, while the centers $C_1$ and $C_2$ of the two primaries are located at $(-\mu, 0)$ and $(1-\mu,0)$, respectively. The total time-independent potential is
\begin{equation}
V(x,y) = - \frac{\mu}{r_2} - \frac{(1 - \mu)}{r_1} - \frac{1}{2}\left( x^2  + y^2 \right),
\label{pot}
\end{equation}
where
\begin{eqnarray}
r_1 &=& \sqrt{\left(x + \mu\right)^2 + y^2},  \nonumber \\
r_2 &=& \sqrt{\left(x + \mu - 1\right)^2 + y^2},
\label{dist}
\end{eqnarray}
are the distances to the respective primaries.

The scaled equations of motion describing the motion of the test body in the corotating frame read
\begin{eqnarray}
\ddot{x} &=& 2\dot{y} - \frac{\partial V(x,y)}{\partial x},  \nonumber \\
\ddot{y} &=& - 2\dot{x} - \frac{\partial V(x,y)}{\partial y}.
\label{eqmot}
\end{eqnarray}
We observe that the equations of motion (\ref{eqmot}) are invariant under the symmetry operation $\Sigma: (x,y,t) \rightarrow (x,-y,-t)$, while for the case $\mu = 1/2$ there is another type of symmetry $\Sigma': (x,y,t) \rightarrow (-x,-y,-t)$. It should be pointed out that these two are the only known independent symmetries regarding the equations of motion of the PCRTBP.

The dynamical system (\ref{eqmot}) admits the well know Jacobi integral
\begin{equation}
J(x,y,\dot{x},\dot{y}) = \frac{1}{2} \left(\dot{x}^2 + \dot{y}^2 \right) + V(x,y) = E,
\label{ham}
\end{equation}
where $\dot{x}$ and $\dot{y}$ are the momenta\footnote{It should be emphasized that the momenta $\dot{x}$ and $\dot{y}$ are not the canonical momenta $p_x$ and $p_y$, respectively.} per unit mass, conjugate to $x$ and $y$, respectively, while $E$ is the numerical value of the energy which is conserved and defines a three-dimensional invariant manifold in the total four-dimensional phase space. Thus, an orbit with a given value of it's energy integral is restricted in its motion to regions in which $E \leq V(x,y)$, while all other regions are forbidden to the test body. It is widely believed that $J$ is the only independent integral of motion for the PCRTBP system [\citealp{P93}]. The energy value $E$ is related with the Jacobi constant by $C = - 2E$.

\begin{figure}[!tH]
\resizebox{\hsize}{!}{\includegraphics{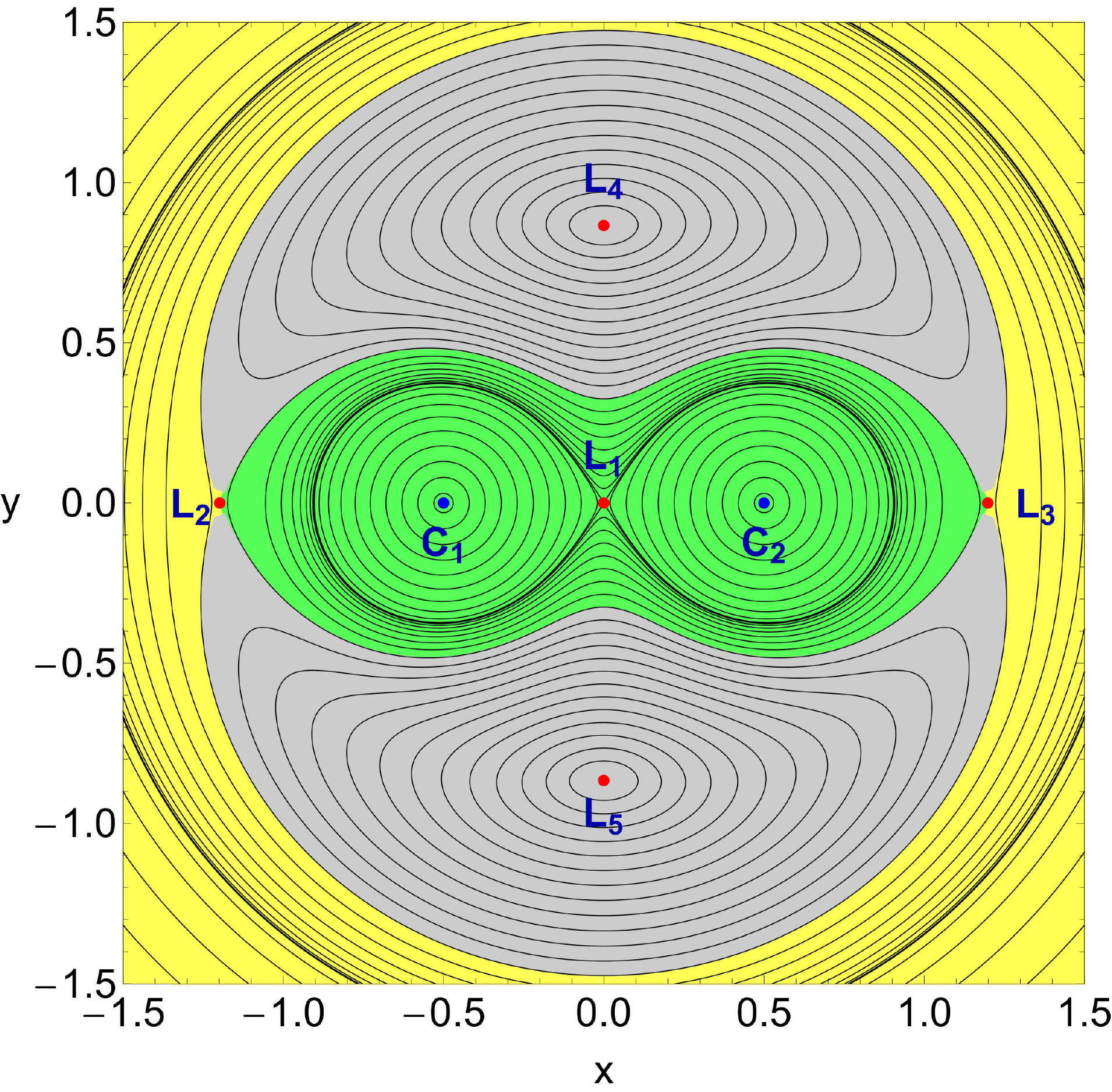}}
\caption{The isolines contours of the constant potential, the location of the centers of the two primaries (blue) and the position of the five Lagrangian points (red), for $\mu = 1/2$. The interior region is indicated in green, the exterior region is shown in yellow, while the forbidden regions of motion are marked with grey.}
\label{conts}
\end{figure}

The dynamical system has five equilibria known as Lagrangian points [\citealp{S67}] at which
\begin{equation}
\frac{\partial V(x,y)}{\partial x} = \frac{\partial V(x,y)}{\partial y} = 0.
\label{lps}
\end{equation}
The isolines contours of constant potential, the position of the five Lagrangian points $L_i, \ i = {1,5}$, as well as the centers of the two primaries are shown in Fig. \ref{conts} where $\mu = 1/2$. Three of them, $L_1$, $L_2$, and $L_3$, are collinear points located in the $x$-axis. The central stationary point $L_1$ at $(x,y) = (0,0)$ is a local minimum of the potential $V(x,y)$. At the other four Lagrangian points it is possible for the test body to move in a circular orbit, while appearing to be stationary in the rotating frame. For this circular orbit, the centrifugal and the gravitational forces precisely balance. The stationary points $L_2$ and $L_3$ at $(x,y) = (\pm r_L,0)$ are saddle points, where $r_L$ is called Lagrangian radius. Let $L_2$ located at $x = -r_L$, while $L_3$ be at $x = +r_L$. The points $L_4$ and $L_5$ on the other hand, are local maxima of the gravitational potential, enclosed by the banana-shaped isolines. The annulus bounded by the circles through $L_2$, $L_3$ and $L_4$, $L_5$ is known as the ``region of coroation" (see also [\citealp{BT08}]). The projection of the four-dimensional phase space onto the physical (or position) space $(x,y)$ is called the Hill's regions and is divided into three domains shown in Fig. \ref{conts} with different colors: (i) the interior region (green) for $-r_L \leq x \leq +r_L$; (ii) the exterior region (yellow) for $x < -r_L$ and $x > r_L$; (iii) the forbidden regions (gray). The boundaries of these Hill's regions are called Zero Velocity Curves (ZVCs) because they are the locus in the physical $(x,y)$ space where the kinetic energy vanishes.

\begin{figure*}[!tH]
\centering
\resizebox{0.8\hsize}{!}{\includegraphics{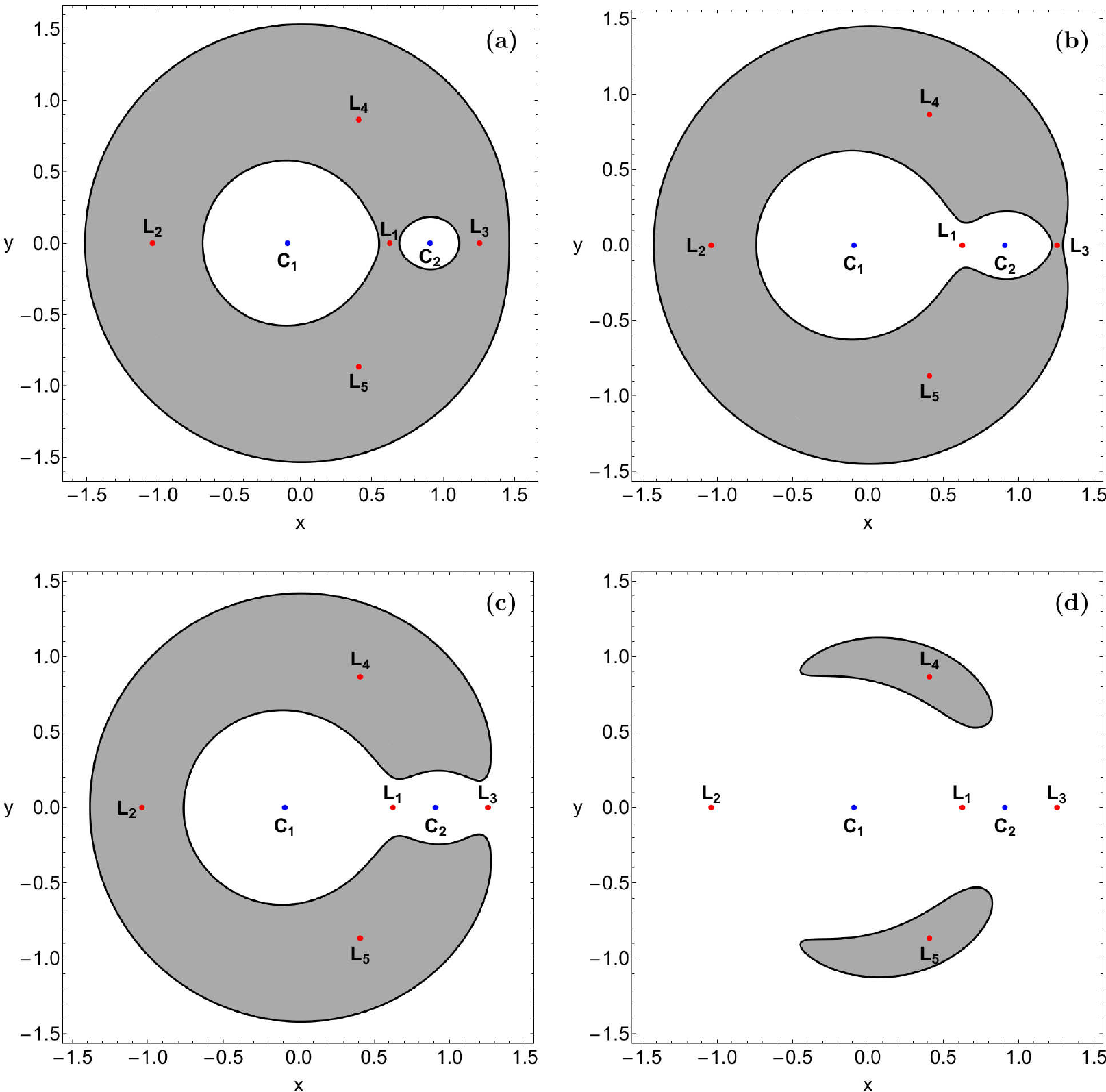}}
\caption{Four possible Hill's region configurations for the PCRTBP system when $\mu = 1/11$. The white domains correspond to the Hill's region, gray shaded domains indicate the forbidden regions, while the thick black lines depict the Zero Velocity Curves (ZVCs). The red dots pinpoint the position of the Lagrangian points, while the positions of the centers of the two primaries are indicated by blue dots. (a-upper left): $E = -1.82$. No transit orbits among the three regions are possible; (b-upper right): $E = -1.73$. The neck around $L_1$ is open; (c-lower left): $E = -1.70$. Both necks around $L_1$ and $L_3$ are open; (d-lower right): $E = -1.50$. The neck around $L_2$ opens, while the forbidden regions diminish. }
\label{isos}
\end{figure*}

The values of the Jacobi integral at the five Lagrangian points $L_i$ are critical energy levels and are denoted as $E_i$ (Note that $E_4 = E_5$). The structure of the equipotential surfaces strongly depends on the value of the energy. In particular, there are five distinct cases
\begin{itemize}
  \item $E < E_1$: All necks are closed therefore, we have only collisional and bounded motion, while transit orbits around the two primaries are not possible.
  \item $E_1 < E < E_3$: Only the neck around $L_1$ is open, so the realms around the two primaries are connected by transit orbits through the open neck.
  \item $E_3 < E < E_2$: The neck around $L_3$ acts as escape channel allowing orbits to escape from the system.
  \item $E_2 < E < E_4$: The necks around both $L_2$ and $L_3$ are open and two symmetrical, with respect to the $y = 0$ axis, forbidden regions are present.
  \item $E > E_4$: The banana-shaped forbidden regions disappear and therefore, motion over the entire physical $(x,y)$ plane is possible.
\end{itemize}
In Fig. \ref{isos}(a-d) we present for $\mu = 1/11$ a characteristic equipotential surface for the first four possible Hill's region configurations. We observe in Fig. \ref{isos}d the two openings (exit channels) at the Lagrangian points $L_2$ and $L_3$ through which the body can leak out. In fact, we may say that these two exits act as hoses connecting the interior region of the system where $-r_L < x < r_L$ with the ``outside world" of the exterior region.

\section{Computational methods and criteria}
\label{cometh}

The motion of the test third body is restricted to a three-dimensional surface $E = const$, due to the existence of the Jacobi integral. With polar coordinates $(r,\phi)$ in the center of the mass system of the corotating frame the condition $\dot{r} = 0$ defines a two-dimensional surface of section, with two disjoint parts $\dot{\phi} < 0$ and $\dot{\phi} > 0$. Each of these two parts has a unique projection onto the configuration physical $(x,y)$ space. Our investigation takes place in both types of projection for a better understanding of the orbital dynamics. In order to explore the behavior of test particles in the PCRTBP system, we need to define samples of initial conditions of orbits whose properties will be identified. For this purpose we define for several values of the total orbital energy $E$, dense uniform grids of $1024 \times 1024$ initial conditions regularly distributed on the $(x,y)$ plane inside the area allowed by the value of the energy.

\begin{figure}[!tH]
\centering
\resizebox{\hsize}{!}{\includegraphics{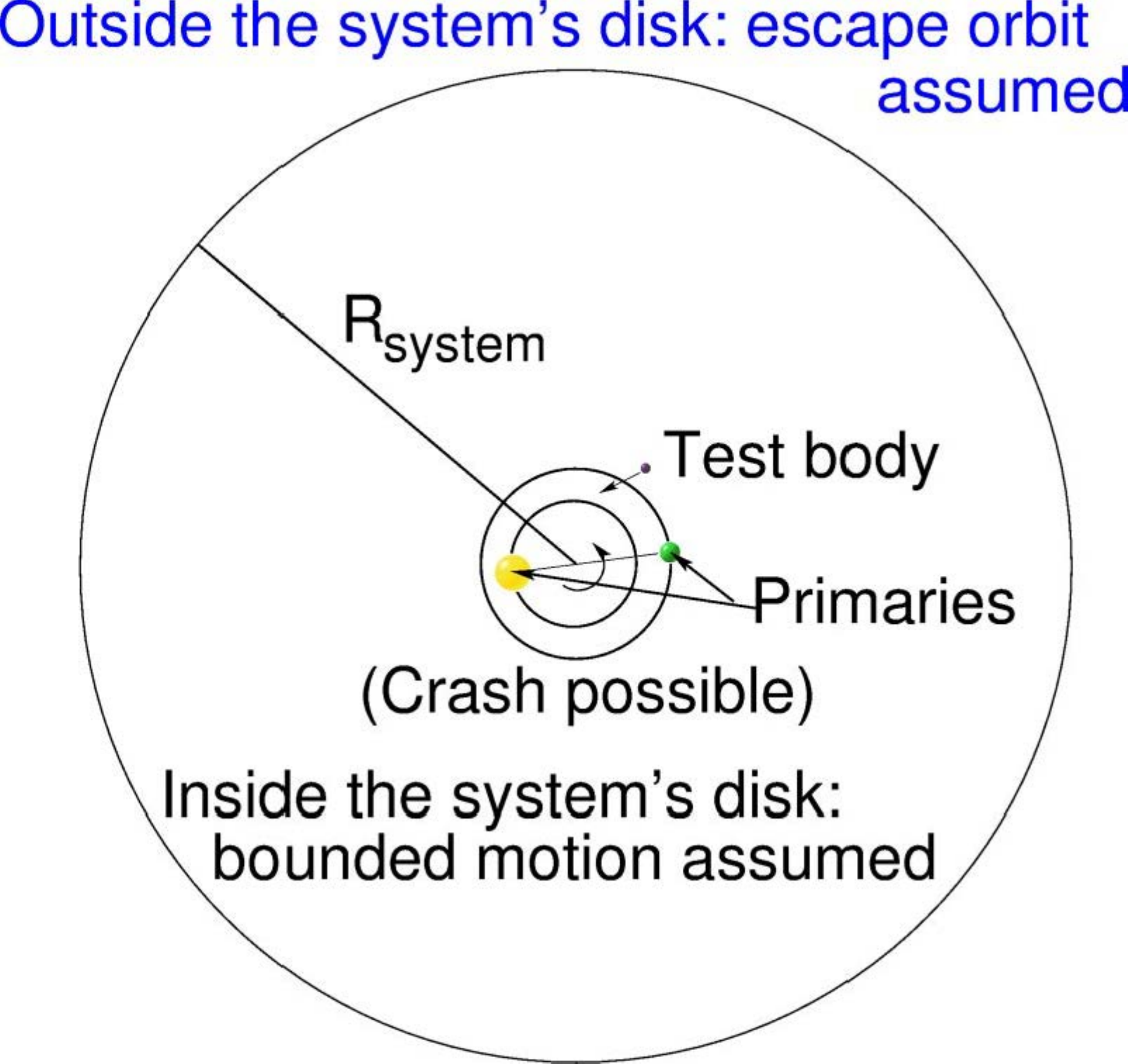}}
\caption{Schematic picture of the three different types of motion. The motion is considered to be bounded if the test body stays confined for integration time $t_{\rm max}$ inside the system's disk with radius $R_d = 10$, while the motion is unbounded and the numerical integration stops when the test body crosses the system's disk with velocity pointing outwards. Collision with one of the primaries occurs when the test body crosses the disk of radius $R_{m1} = 10^{-4}$ and $R_{m2} = R_{m1} \times (2 \mu)^{1/3}$ of one of the primaries.}
\label{crit}
\end{figure}

In the PCRTBP system the configuration space extends to infinity thus making the identification of the type of motion of the test body for specific initial conditions a rather demanding task. There are three possible types of motion for the test body: (i) bounded motion around one of the primaries, or even around both; (ii) escape to infinity; (iii) collide to one of the primaries. Now we need to define appropriate numerical criteria for distinguishing between these three types of motion. The motion is considered as bounded if the test body stays confined for integration time $t_{\rm max}$ inside the system's disk with radius $R_d$ and center coinciding with the center of mass origin at $(0,0)$. Obviously, the higher the values of $t_{\rm max}$ and $R_d$ the more plausible becomes the definition of bounded motion and in the limit $t_{\rm max} \rightarrow \infty$ the definition is the precise description of bounded motion in a finite disk of radius $R_d$. Consequently, the higher these two values, the longer the numerical integration of initial conditions of orbits lasts. In our calculations we choose $t_{\rm max} = 10^4$ and $R_d = 10$ (see Fig. \ref{crit}). It should be emphasized that for low values of $t_{\rm max}$ the fractal boundaries of stability islands corresponding to bounded motion become more smooth. Moreover, an orbit is identified as escaping and the numerical integration stops if the test body body intersects the system's disk with velocity pointing outwards at a time $t_{\rm esc} < t_{\rm max}$. Finally, a collision with one of the primaries occurs if the test body, assuming it is a point mass, crosses the disk with radius $R_m$ around the primary, where in our case we choose $R_{m1} = 10^{-4}$ and $R_{m2} = R_{m1} \times (2 \mu)^{1/3}$. Here is should be noted that in generally it is assumed that the radius of a celestial body (e.g., a planet) is directly proportional to the cubic root of its mass.

\begin{figure}[!tH]
\centering
\resizebox{\hsize}{!}{\includegraphics{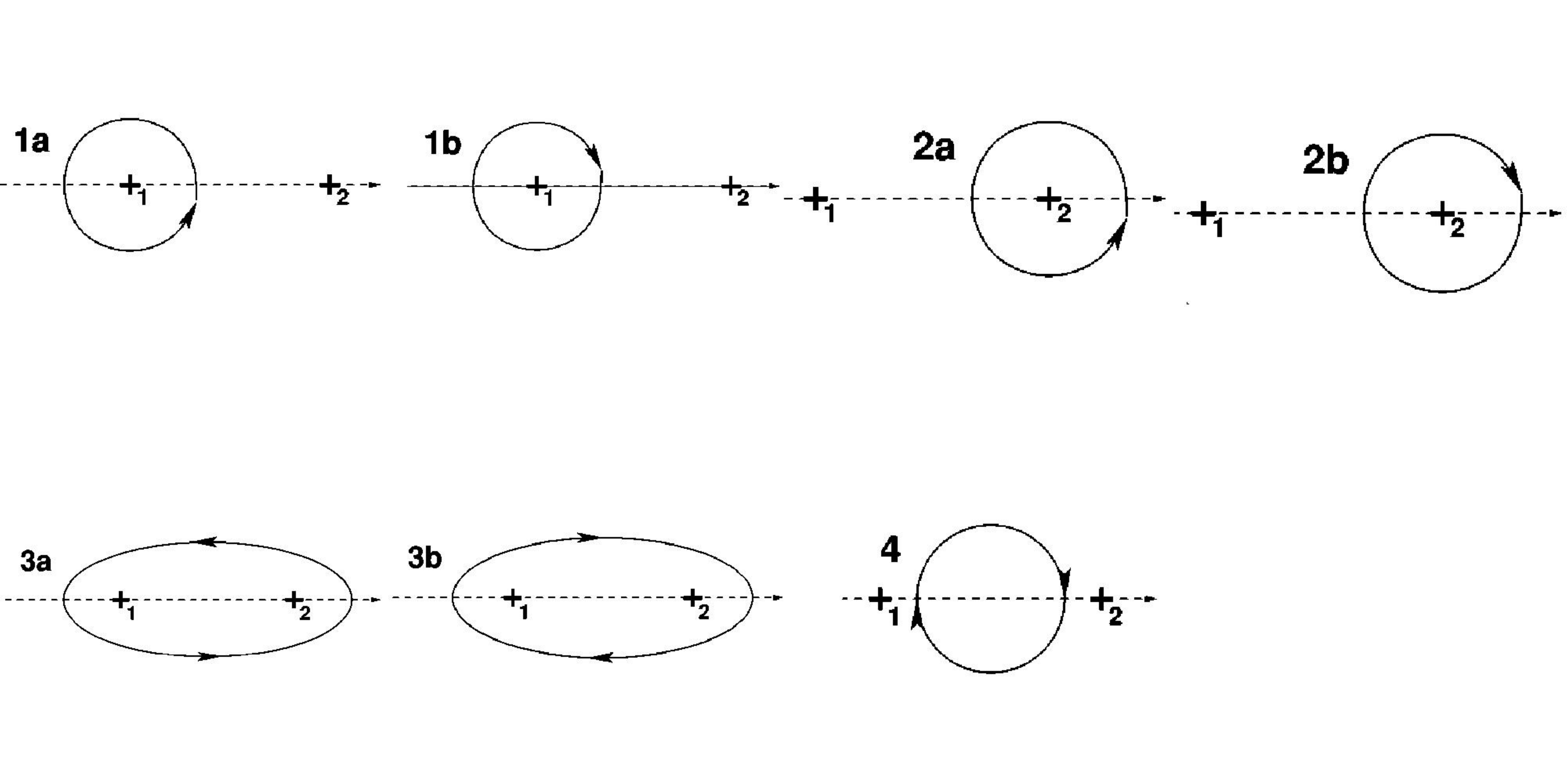}}
\caption{Characteristic orbit examples of the seven main types of regular orbits.}
\label{types}
\end{figure}

The vast majority of bounded motion corresponds to initial conditions of regular orbits. It therefore seems appropriate to further classify initial conditions of ordered orbits into regular families. For this task we use the symbolic orbit classification which was also used in Papers I and II. According to this method orbits are classified by taking into account their orientation with respect to the centers of the two primaries $C_1$ and $C_2$, as well as their rotation (clockwise or counterclockwise). In particular, the orbit classification is based on an automatic detection of $x$ axis passages of the test body. Furthermore, two consecutive $x$ axis passages define a half rotation with respect to the fixed centers of the two primary bodies. In Fig. \ref{types} we present characteristic orbit examples of the seven main types of regular orbits.

As it was stated earlier, in our computations, we set $10^4$ time units as a maximum time of numerical integration. The vast majority of escaping orbits (regular and chaotic) however, need considerable less time to escape from the system (obviously, the numerical integration is effectively ended when an orbit moves outside the system's disk and escapes). Nevertheless, we decided to use such a vast integration time just to be sure that all orbits have enough time in order to escape. Remember, that there are the so called ``sticky orbits" which behave as regular ones during long periods of time. Here we should clarify, that orbits which do not escape after a numerical integration of $10^4$ time units are considered as non-escaping or trapped.

The equations of motion (\ref{eqmot}) for the initial conditions of all orbits are forwarded integrated using a double precision Bulirsch-Stoer \verb!FORTRAN 77! algorithm (e.g., [\citealp{PTVF92}]) with a small time step of order of $10^{-2}$, which is sufficient enough for the desired accuracy of our computations. Here we should emphasize, that our previous numerical experience suggests that the Bulirsch-Stoer integrator is both faster and more accurate than a double precision Runge-Kutta-Fehlberg algorithm of order 7 with Cash-Karp coefficients. Throughout all our computations, the Jacobian energy integral (Eq. (\ref{ham})) was conserved better than one part in $10^{-11}$, although for most orbits it was better than one part in $10^{-12}$. For collisional orbits where the test body moves inside a region of radius $10^{-2}$ around one of the primaries the Lemaitre's global regularization method is applied.

\section{Numerical results \& Orbit classification}
\label{numres}

The main objective of our investigation is to classify initial condition of orbits in the physical $(x,y)$ plane into three categories: (i) bounded orbits; (ii) escaping orbits and (iii) collisional orbits, distinguishing simultaneously regular orbits into different types. Furthermore, two additional properties of the orbits will be examined: (i) the time-scale of collison and (ii) the time-scale of the escapes (we shall also use the terms escape period or escape rates). In the present paper, we shall explore these dynamical quantities for various values of the total orbital energy, as well as for the mass ratio $\mu$. In the following color-coded grids (or orbit type diagrams - OTDs) each pixel is assigned a color according to the orbit type. Thus the initial conditions of orbits are classified into bounded motion of a few types, unbounded escaping motion and collisional motion. In this special type of Poincar\'{e} surface of section the phase space emerges as a close and compact mix of escape basins, collision basins and stability islands. For each case we examine three energy levels that correspond to the last three Hill's regions configurations explained earlier in Fig. \ref{conts}. The cases $E < E_1$ and $E_1 < E < E_3$ are not very interesting because they contain only collisional and bounded motion, so we decided not to include them.

\subsection{Case I: The Jefferys system $(\mu = 1/3)$}
\label{cas2}

\begin{figure*}[!tH]
\centering
\resizebox{\hsize}{!}{\includegraphics{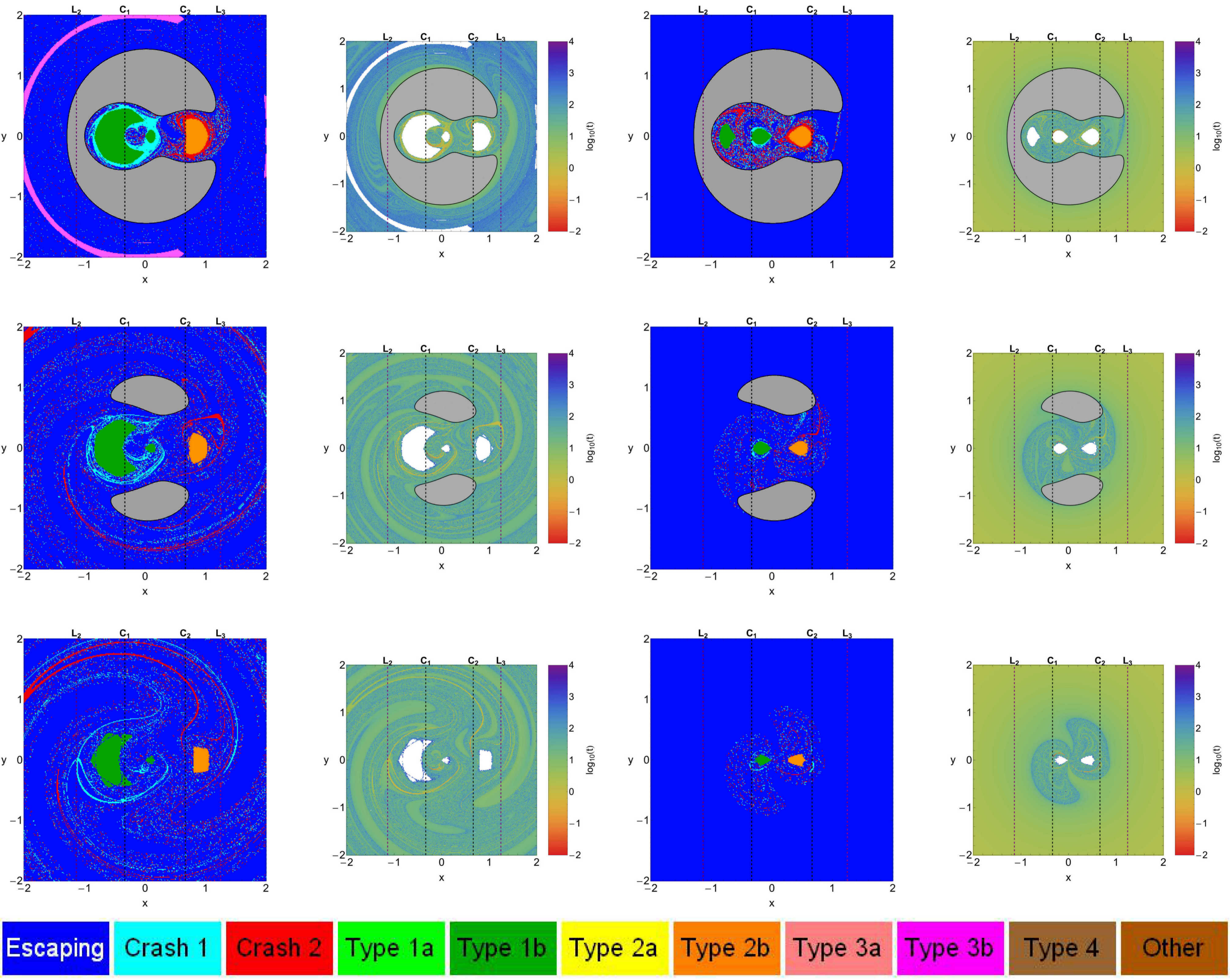}}
\caption{The orbital structure of the physical $(x,y)$ plane in a corotating frame of reference is given using orbit type diagrams (OTDs) for three energy levels and for both parts $\dot{\phi} < 0$ (first column) and $\dot{\phi} > 0$ (third column) of the surface of section $\dot{r} = 0$, when $\mu = 1/3$. (Top row): $E = -1.70$; (middle row): $E = -1.50$; (bottom row): $E = -1.30$. The vertical black dashed lines denote the centers of the two primaries, wile the vertical purple dashed lines indicate the position of the Lagrangian points $L_2$ and $L_3$. The color bar contains the color code which relates the types of orbits presented in Fig. \ref{types} with different colors. (second and fourth columns): Distribution of the corresponding escape and collisional time of the orbits on the two parts of the physical $(x,y)$ space. The darker the color, the larger the escape/collision time. Initial conditions of bounded regular orbits are shown in white.}
\label{grd2}
\end{figure*}

Our exploration begins considering the case where the mass ratio is $\mu = 1/3$ (known also as the Jefferys system [\citealp{J71})]. In Fig. \ref{grd2} the OTD decompositions for both $\dot{\phi} < 0$ (first column) and $\dot{\phi} > 0$ (third column) reveal the structure of the physical $(x,y)$ space for three energy levels, where the several types of orbits are indicated with different colors. The color code is explained in the color bar at the bottom of the figure. The black solid lines in the two types of plots denote the Zero Velocity Curve, while the inaccessible forbidden regions are marked in gray. The color of a point represents the orbit type of a test body which has been launched with pericenter position at $(x,y)$. It should be emphasized that for $\mu \neq 1/2$ the origin does not coincide with the Lagrangian point $L_1$, while the equations of motion do not admit the symmetry $\Sigma'$. When $E = -1.70$ we observe that in both cases the interior region is filled with three types of initial conditions of orbits: (i) regular, (ii) collide to one of the primaries and (iii) escaping orbits. Regular motion dominates the interior region and large stability islands are shown near the centers of the primaries. These stability islands contain quasi-periodic orbits which are symmetrical with respect to a reflection over the $x$ axis and they move in clockwise sense, hence, retrograde in relation to the rotating system of coordinates. Moreover, the stability regions are surrounded by a chaotic mix of domains of collisional orbits with respect to the first and the second primary body. On the other hand, the exterior region contains mostly initial conditions of escaping orbits however, in the $\dot{\phi} < 0$ plot we have to point out the existence of an open stability ring containing initial conditions of regular orbits that circulate clockwise around both primaries. As the value of the energy increases and the area of forbidden region is reduced it is seen that the stability ring disappears, while in both the interior and exterior regions basins of escaping and collisional orbits are formed. For $E = -1.30 > E_4$ the test body has full access to the entire physical $(x,y)$ plane. Two regions of bounded motion are shown around the primaries, where for each stability island there is a parent periodic orbit at its center. Furthermore, due to the rotation of the primaries the collision basins wind out in spiral form in the outer regions of the $\dot{\phi} < 0$ plot. Crash basins are also present in the immediate vicinity of the origin. At this point we would like to stress out that the area of collision basins is several orders of magnitude larger than the total size of the primary body's disks. In the OTD for $\dot{\phi} > 0$ the escape basin cover the vast majority of the configuration space, while the total size of stability regions is less than for $\dot{\phi} < 0$. It is interesting to note that all regular orbits in all three energy levels were found to be retrograde thus traveling in clockwise sense. We observe that the area of regular motion around primary 1 is fragmented. Due to the smaller mass of primary 2 the stability area with respect to primary 2 is smaller than the total size of bounded motion around primary 1. It should be noted that the basins containing orbits that collide to primary 2 are much smaller, with respect to the collision basins of primary 1, and they appear only as thin filaments. For both parts of the configuration space the total area occupied by initial conditions corresponding to bounded motion around primary 1 and primary 2 decreases with increasing energy.

The second and third columns of Fig. \ref{grd2} show how the corresponding escape and collision times of orbits are distributed on the two parts of the physical $(x,y)$ space. Light reddish colors correspond to fast escaping/collisional orbits, dark blue/purpe colors indicate large escape/collision rates, while white color denote stability islands of regular motion. Note that the scale on the color bar is logarithmic. Inspecting the spatial distribution of various different ranges of escape time, we are able to associate medium escape time with the stable manifold of a non-attracting chaotic invariant set, which is spread out throughout this region of the chaotic sea, while the largest escape time values on the other hand, are linked with sticky motion around the stability islands of the two primaries. As for the collision time we see that orbits with initial conditions inside the collision basins collide with one of the primaries almost immediately. It is seen, that orbits with initial conditions inside the escape and collision basins have the smallest escape/collision rates, while on the other hand, the longest escape/collision times correspond to orbits with initial conditions in the fractal regions of the plots. At this point, we would like to point out that the basins of escape can be easily distinguished being the regions with intermediate greenish colors indicating fast escaping orbits. Indeed, our numerical calculations suggest that orbits with initial conditions inside these basins need no more than 50 time units to escape from the system. Furthermore, the collision basins are shown with reddish colors where the corresponding collision time is less than one time unit.

\subsection{Case II: The Moulton system $(\mu = 1/5)$}
\label{cas3}

\begin{figure*}[!tH]
\centering
\resizebox{\hsize}{!}{\includegraphics{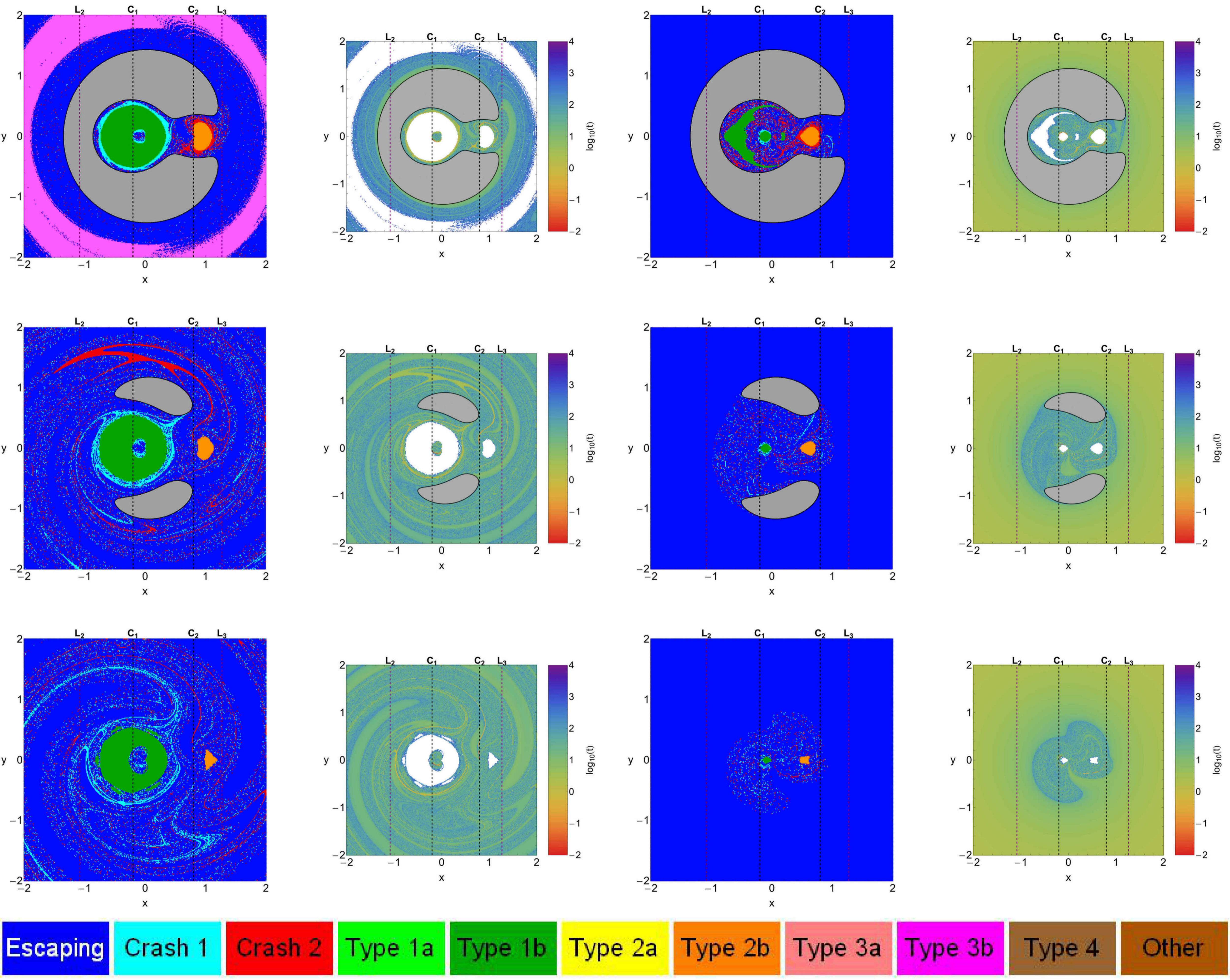}}
\caption{The orbital structure of the physical $(x,y)$ plane in a corotating frame of reference is given using orbit type diagrams (OTDs) for three energy levels and for both parts $\dot{\phi} < 0$ (first column) and $\dot{\phi} > 0$ (third column) of the surface of section $\dot{r} = 0$, when $\mu = 1/5$. (Top row): $E = -1.70$; (middle row): $E = -1.50$; (bottom row): $E = -1.30$. (second and fourth columns): Distribution of the corresponding escape and collisional time of the orbits on the two parts of the physical $(x,y)$ space.}
\label{grd3}
\end{figure*}

The next case under investigation involves the Moulton system [\citealp{M20}], that is when $\mu = 1/5$. Again, all the different aspects of the numerical approach remain exactly the same as in the two previously studied cases. The orbital structure of the physical $(x,y)$ plane through the OTD decompositions for both $\dot{\phi} < 0$ (first column) and $\dot{\phi} > 0$ (third column) parts of the configuration space distinguishing between the main types of orbits for three energy levels is shown in Fig. \ref{grd3}. It is seen that for the $\dot{\phi} < 0$ part of the configuration space and for $E = -1.70$ the outside stability ring is closed again thus dividing the exterior region into two realms. The stability region around primary 1 resembles a section of a torus, where the central hole contains a highly fractal mixture of initial conditions of collisional and escaping orbits. This central fractal hole survives for all mass ratios $\mu < 1.5$, even though too tiny to be visible. Furthermore, one may observes that the collision basins are mainly located around the two stability islands. In particular, at the boundaries of the stability island around primary 1 we see a thin ring of initial conditions of orbits that collide to primary 1. When $E = -1.50$ one may identify the complicated spiral structure of the basin containing the initial conditions of orbits that collide to primary 2. The $\dot{\phi} > 0$ part of the configuration space displays, once more, the same structure as in the previous two cases, where we found that the vast majority of the integrated initial conditions correspond to orbits which escape from the system. Our numerical calculations indicate that in both parts of the configuration space the extent of the stability islands is reduced as we proceed to higher energy levels. The second and third columns of Fig. \ref{grd3} show how the corresponding escape and collision times of orbits are distributed on the two parts of the physical $(x,y)$ space.

\subsection{Case III: The Darwin system $(\mu = 1/11)$}
\label{cas4}

\begin{figure*}[!tH]
\centering
\resizebox{\hsize}{!}{\includegraphics{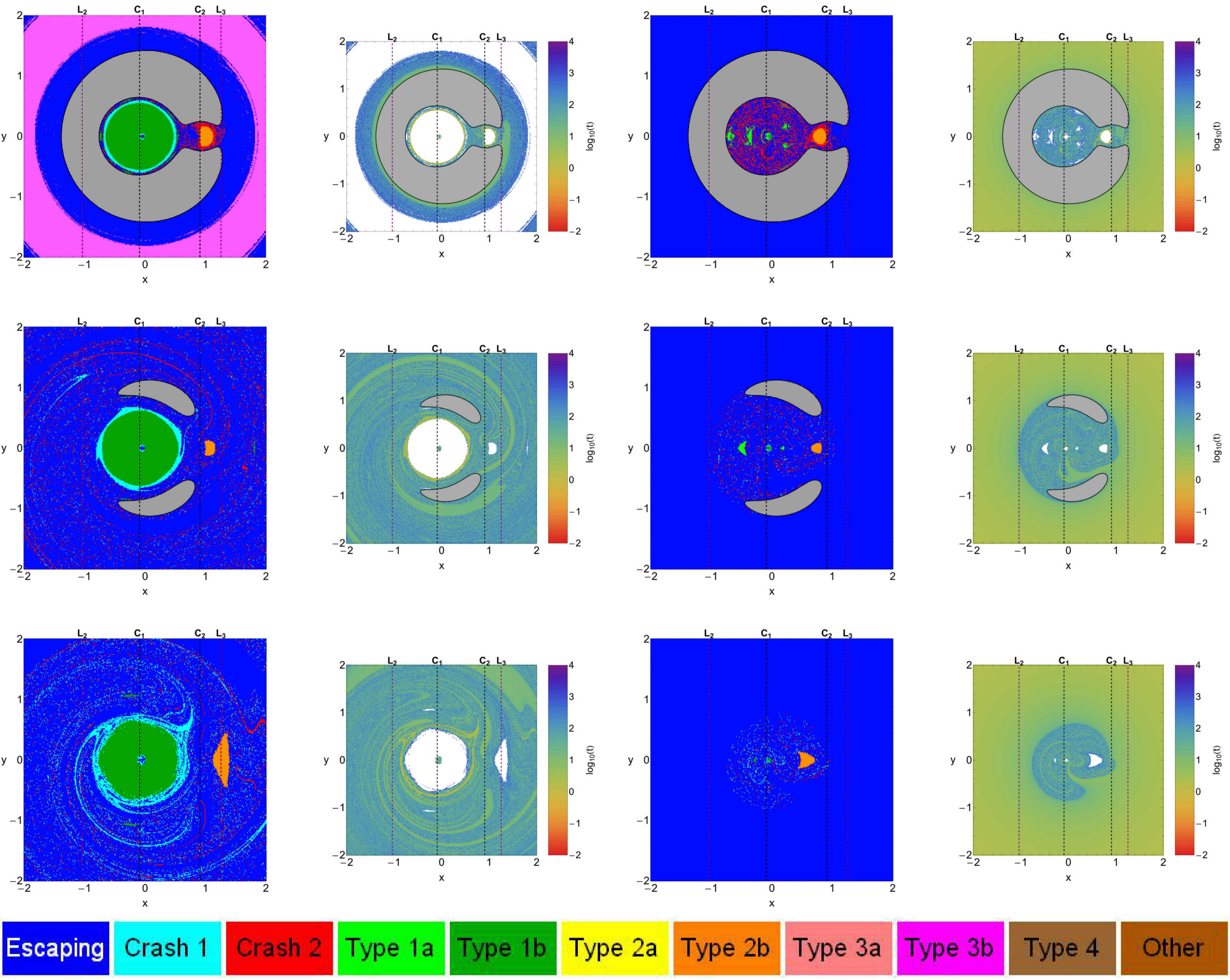}}
\caption{The orbital structure of the physical $(x,y)$ plane in a corotating frame of reference is given using orbit type diagrams (OTDs) for three energy levels and for both parts $\dot{\phi} < 0$ (first column) and $\dot{\phi} > 0$ (third column) of the surface of section $\dot{r} = 0$, when $\mu = 1/11$. (Top row): $E = -1.70$; (middle row): $E = -1.50$; (bottom row): $E = -1.30$. (second and fourth columns): Distribution of the corresponding escape and collisional time of the orbits on the two parts of the physical $(x,y)$ space.}
\label{grd4}
\end{figure*}

Our exploration continues with the Darwin system [\citealp{D97}], that is when $\mu = 1/11$, following once more the same numerical methods. The orbital structure of the physical $(x,y)$ plane through the OTD decompositions for both $\dot{\phi} < 0$ (first column) and $\dot{\phi} > 0$ (third column) parts of the configuration space distinguishing between the main types of orbits for three energy levels is shown in Fig. \ref{grd4}. We observe that for $E = -1.70$ the outside ring containing initial conditions of orbits that circulate around both primaries is closed and more thick with respect to that seen in the previous cases. The extent of the stability island of initial conditions of orbits that circulate around primary 1 seems to be unaffected by the increasing energy. The island of bounded motion around primary 2 on the other hand, decreases in size for $E = -1.50$, while for $E = -1.30 > E_4$ it occupies more area on the physical space than for $E = -1.70$. Moreover, we see that for $E > -1.70$ the collision basin around the stability island of primary body 2 dissolves and initial conditions of orbits that collide to primary 2 produce only thin filaments in the configuration space. Initial conditions of orbits that lead to escape cove, once more, the vast majority of the $\dot{\phi} > 0$ part of the configuration space. For all tested energy levels we found that bounded motion around primary 2 corresponds to a single stability islands, while on the contrary bounded motion around primary body 1 correspond to several small stability islands which are embedded in the fractal region of the $(x,y)$ plane. The second and third columns of Fig. \ref{grd4} show how the corresponding escape and collision times of orbits are distributed on the two parts of the physical $(x,y)$ space.

\subsection{Case IV: The Earth-Moon system $(\mu = 1/82.3)$}
\label{cas5}

\begin{figure*}[!tH]
\centering
\resizebox{\hsize}{!}{\includegraphics{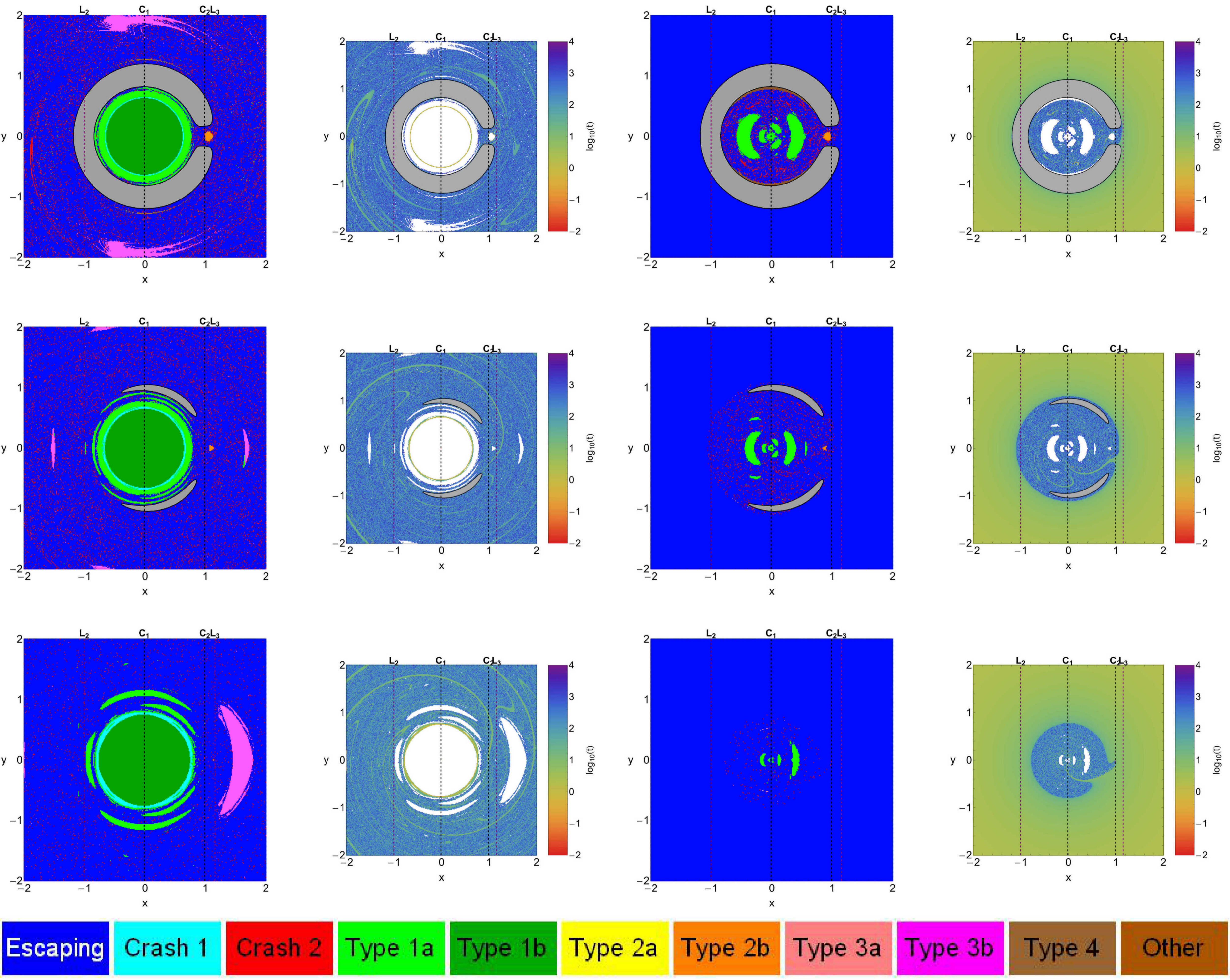}}
\caption{The orbital structure of the physical $(x,y)$ plane in a corotating frame of reference is given using orbit type diagrams (OTDs) for three energy levels and for both parts $\dot{\phi} < 0$ (first column) and $\dot{\phi} > 0$ (third column) of the surface of section $\dot{r} = 0$, when $\mu = 1/82.3$. (Top row): $E = -1.55$; (middle row): $E = -1.50$; (bottom row): $E = -1.30$. (second and fourth columns): Distribution of the corresponding escape and collisional time of the orbits on the two parts of the physical $(x,y)$ space.}
\label{grd5}
\end{figure*}

The last case concerns the Earth-Moon system where $\mu = 1/82.3$. In the following Fig. \ref{grd5} we present the orbital structure of the physical $(x,y)$ plane through the OTD decompositions for both $\dot{\phi} < 0$ (first column) and $\dot{\phi} > 0$ (third column) parts of the configuration space distinguishing between the main types of orbits for three energy levels. For $E = -1.55$ we see that the shape of the stability region corresponding to type 3b regular orbits has changed. The ring annulus is no linger present thus giving its place to two islands which are symmetrical to the $y = 0$ axis. We also observe that the central collision basin around primary 1 separates stable motion of type 1a (counterclockwise) from bounded motion of type 1b (clockwise). With increasing energy the size of the stability regions around primary 1 slightly increases while on the other hand, the area on the physical plane occupied by initial conditions corresponding to bounded motion around primary 2 constantly decreases and for $E = -1.30$ is hardly identified. For $E = -1.50$ the stability islands of the type 3b regular orbits change position in the $(x,y)$ plane, while for $E = -1.30$ a relatively extended type 3b stability island emerges. The $\dot{\phi} > 0$ part of the configuration space displays a very interesting orbital structure. It is seen, that only orbits of type 1a exist which means that only regular orbits which move clockwise are present. Moreover, there are several stability islands of type 1b orbits instead of one single island like in the $\dot{\phi} < 0$ part of the configuration space. Finally, it should be emphasized that since the third test body is launched perpendicularly to the radius vector (i.e., $\dot{r} = 0$), collide to primary body 1 completely disappears. The second and third columns of Fig. \ref{grd5} show how the corresponding escape and collision times of orbits are distributed on the two parts of the physical $(x,y)$ space.

\begin{figure*}[!tH]
\centering
\resizebox{\hsize}{!}{\includegraphics{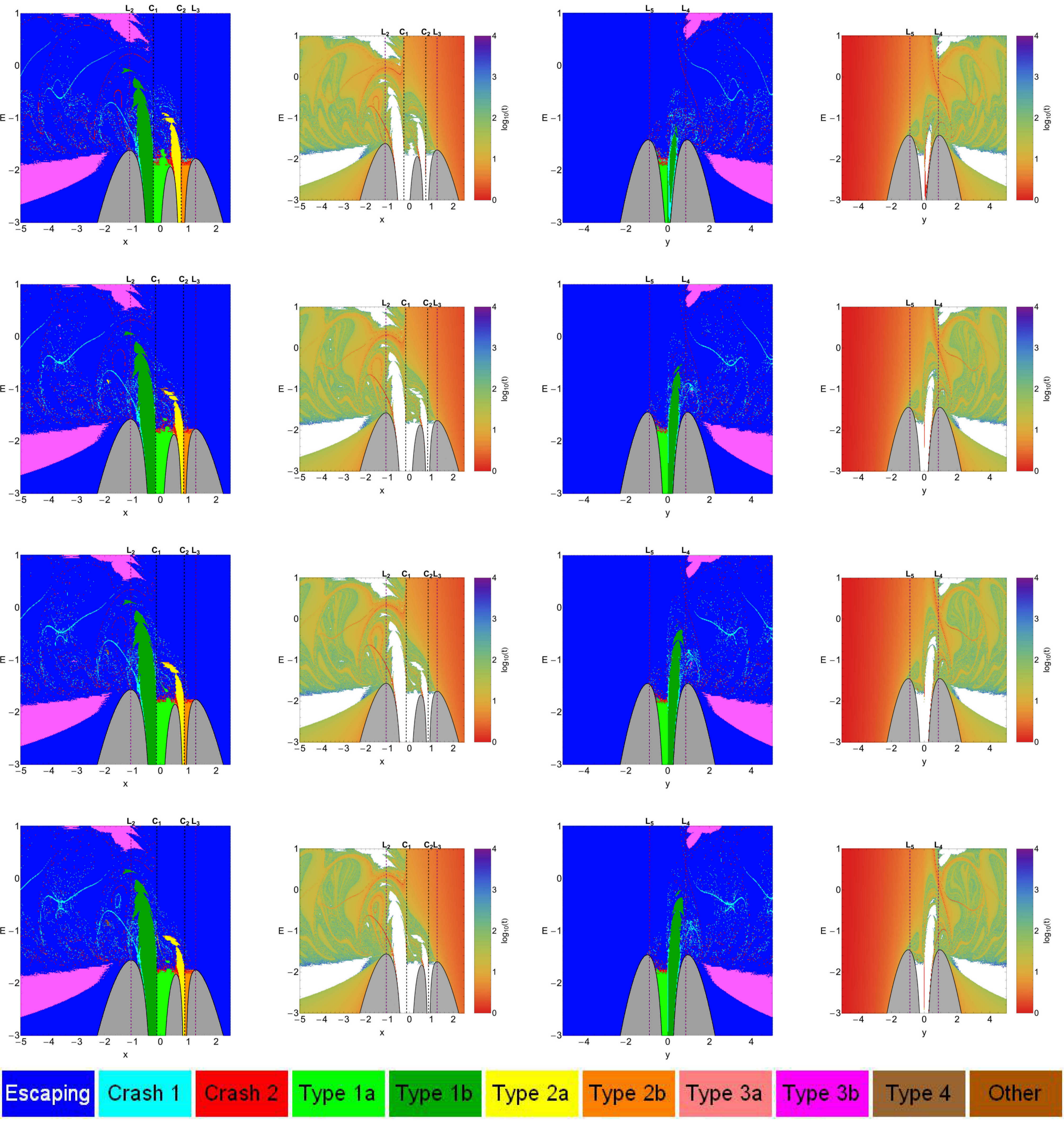}}
\caption{Orbital structure of the (first column): $(x,E)$ plane; (third column): $(y,E)$ plane when (first row): $\mu = 1/4$; (second row): $\mu = 1/6$; (third row): $\mu = 1/7$ and (fourth row): $\mu = 1/8$. The vertical black dashed lines denote the centers of the two primaries, wile the vertical purple dashed lines indicate the position of the Lagrangian points $L_2$ to $L_5$. The color bar contains the color code which relates the types of orbits presented in Fig. \ref{types} with different colors. (second and fourth columns): The distribution of the corresponding escape/collisional times of the orbits.}
\label{xyEt}
\end{figure*}

\begin{figure*}[!tH]
\centering
\resizebox{0.6\hsize}{!}{\includegraphics{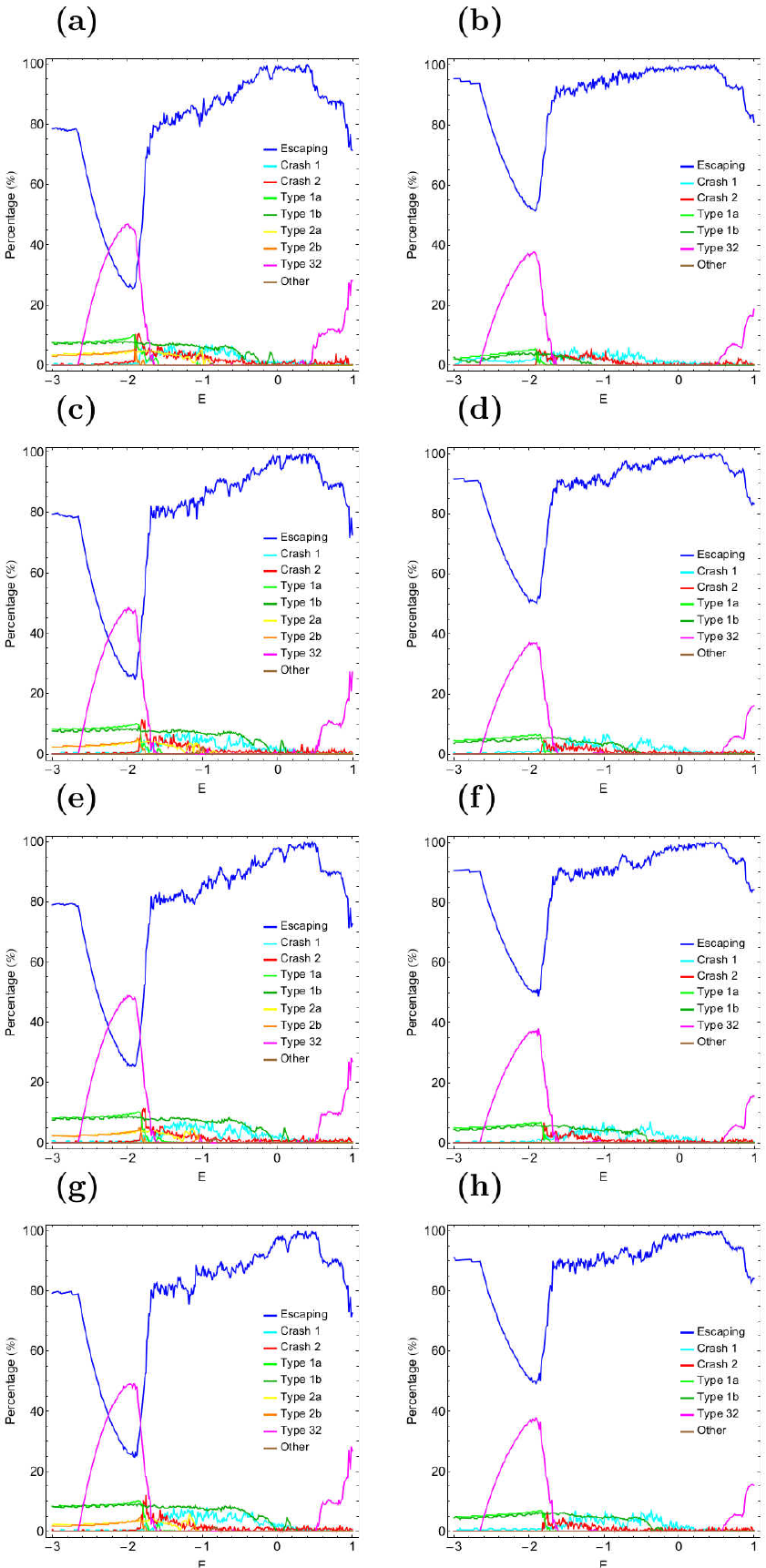}}
\caption{Evolution of the percentages of all types of orbits in the (left column): $(x,E)$ planes and (right column): $(y,E)$ planes shown in Fig. \ref{xyEt}. (first row): $\mu = 1/4$; (second row): $\mu = 1/6$; (third row): $\mu = 1/7$ and (fourth row): $\mu = 1/8$.}
\label{percs}
\end{figure*}

Before closing this section we would like to emphasize that the OTDs given in Figs. \ref{grd2}, to \ref{grd5} have both fractal and non-fractal (smooth) boundary regions which separate the escape basins and the collisional basins. Such fractal basin boundaries is a common phenomenon in leaking Hamiltonian systems (e.g., [\citealp{BGOB88}, \citealp{dML99}, \citealp{dMG02}, \citealp{STN02}, \citealp{ST03}, \citealp{TSPT04}]). In the PCRTBP system the leakages are defined by both escape and collisional conditions thus resulting in three exit modes. However, due to the high complexity of the basin boundaries, it is very difficult, or even impossible, to predict in these regions whether the test body (e.g., a satellite, asteroid, planet etc) collides with a primary body or escapes from the dynamical system.

\section{An overview analysis}
\label{over}

The color-coded OTDs in both parts of the physical $(x,y)$ space provide sufficient information on the phase space mixing however, for only a fixed value of the energy integral and also for orbits that traverse the surface of section either directly (progradely) or retrogradely. H\'{e}non back in the late 60s [\citealp{H69}], introduced a new type of plane which can provide information not only about stability and chaotic regions but also about areas of bounded and unbounded motion using the section $y = \dot{x} = 0$, $\dot{y} > 0$ (see also [\citealp{BBS08}, \citealp{BBS09b}]). In other words, all the orbits of the test particles are launched from the $x$-axis with $x = x_0$, parallel to the $y$-axis $(y = 0)$. Consequently, in contrast to the previously discussed types of planes, only orbits with pericenters on the $x$-axis are included and therefore, the value of the energy $E$ can be used as an ordinate. In this way, we can monitor how the energy influences the overall orbital structure of our dynamical system using a continuous spectrum of energy values rather than few discrete energy levels. In the first column of Fig. \ref{xyEt} we present the orbital structure of the $(x,E)$ plane for four other interesting values of the mass ratio $\mu$ [\citealp{J71}] when $E \in [-3,1]$, while in the second column of the same figure the distribution of the corresponding escape/collision times of orbits is depicted.

We observe the presence of several types of regular orbits around the two primary bodies. Being more precise, on both sides of the primaries we identify stability islands corresponding to both direct (counterclockwise) and retrograde (clockwise) quasi-periodic orbits. It is seen that a large portion of the exterior region, that is for $x < - x_L$ and $x > x_L$, is covered by initial conditions of escaping orbits however, at the left-hand side of the same plane two stability islands of type 3b regular orbits are observed. Additional numerical calculations reveal that for much lower values of $x$ $(x < -5)$ these two stability islands are joined and form a crescent-like shape. Furthermore, orbits with initial conditions very close to vertical lines $x = C_1$ and $x = C_2$, or in other words close to the centers of the primaries collide almost immediately with them. The smaller the mass ratio the closer the center of primary body 1 wanders to the origin. We also see that collision basins leak outside the interior region, mainly outside $L_2$, and create complicated spiral shapes in the exterior region. It should be pointed out that in the blow-ups of the diagram several additional very small islands of stability have been identified\footnote{An infinite number of regions of (stable) quasi-periodic (or small scale chaotic) motion is expected from classical chaos theory.}. We observe that as we proceed to lower mass ratios the area of collisional orbits with respect to primary body 2 shrinks, while on the other hand, the area representing regular motion around primary body 1 significantly grows. Moreover, the extent of the area corresponding to bounded motion around the second primary body decreases. Another interesting phenomenon is the fact that the boundaries between the several stability islands and the escape basins become smoother for a decreasing mass ratio.

In the same vein, we can construct another interesting type of plane following the philosophy of the $(x,E)$ plane. In particular we use the section $x = \dot{y} = 0$, $\dot{x} > 0$, so all the orbits of the test particles are initiated for the $y$-axis having $y = y_0$, parallel to the $x$-axis. Therefore, only orbits with pericenters on the $y$-axis are included and the value of the energy $E$ can be used again as an ordinate. In the third column of Fig. \ref{xyEt} the orbital structure of the $(y,E)$ plane is presented for four the same values of the mass ratio when $E \in [-3,1]$, while the distribution of the corresponding escape/collision times of orbits is given in the last column of the same figure. The structure of this new type of plane present similarities with respect to the $(x,E)$ plane however, there are also major differences. For example, this time the crescent-like stability islands containing initial conditions of type 3b regular orbits is located in the right hand side of the plane. Furthermore, around $y = 0$ we observe the presence of stability islands corresponding to both type 1a and 1b regular orbits but there is no indication whatsoever of stability islands with initial conditions of regular motion around primary body 2 (types 2a and 2b). Another interesting issue is that the collision basins leak outside the interior region, mainly outside $L_4$, and create complicated spiral shapes. it should also pointed out that as the value of the mass ratio decreases the area of the stability motion around primary 1 grows.

It would be of particular interest to monitor how the total orbital energy influences the percentages of all types of orbits. The following Fig. \ref{percs}(a-h) shows the evolution of the percentages of all types of orbits identified in the $(x,E)$ and $(y,E)$ planes of Figs. \ref{xyEt} as a function of the total orbital energy. it is evident that in both types of planes the change on the value of the energy affects mostly the rates of escaping and regular type 3b orbits. In particular we see that in the interval $-2.7 < E < -2$ the percentage of escaping orbits exhibits a drop, while at the same time the rate of regular type 3b orbits increases and for $-2.3 < E < -1.9$ it is the most popular type of orbits in the $(x,E)$ planes. Escaping orbits dominate the two types of planes and especially for $-1.4 < E < 0.6$ they cover more that 90\% of the planes. All the other types of orbits are much less affected by the change on the energy and the corresponding percentages fluctuate (less than 10\%) throughout.

\section{Discussion and conclusions}
\label{disc}

The scope of this work was to shed some light to the properties of motion in the planar circular restricted three-body problem (PCRTBP) and try to classify the orbits into categories. We continued the work initiated in Papers I and II following similar numerical techniques. We managed to distinguish between bounded, escaping and collisional orbits and we also located the basins of escape/collison, finding correlations with the corresponding escape/collision times of the orbits. Our extensive and thorough numerical investigation strongly suggests, that the overall motion of a test body under the gravitational filed of two primaries is a very complicated procedure. Taking into account the extended primary bodies the model is much more applicable to realistic scenarios of celestial systems that the pure PCRTBP. In particular, we considered specific mass ratios that correspond to well-know celestial systems (e.g., Jefferys, Moulton, Darwin and the Earth-Moon system). At this point we would like to emphasize that the cases studied in section \ref{over} regarding the mass ratio are numerically investigated for the first time in this work.

We defined for several values of the total orbital energy $E$, dense uniform grids of $1024 \times 1024$ initial conditions regularly distributed on both parts ($\dot{\phi} < 0$ and $\dot{\phi} > 0$) the physical $(x,y)$ plane inside the area allowed by the value of the energy. For the numerical integration of the orbits in each type of grid, we needed about between 5 hours and 9 days of CPU time on a Pentium Dual-Core 2.2 GHz PC, depending on the escape and collisional rates of orbits in each case. For each initial condition, the maximum time of the numerical integration was set to be equal to $10^4$ time units however, when a particle escaped or collided with one of the primaries the numerical integration was effectively ended and proceeded to the next available initial condition.

The present article provides quantitative information regarding the escape and collision dynamics in the PCRTBP Hamiltonian system. The main numerical results of our research can be summarized as follows:
\begin{enumerate}
 \item In all examined cases, areas of bounded motion and regions of initial conditions leading to escape in a given direction (basins of escape), were found to exist in both parts of the configuration space. The several escape and collisional basins are very intricately interwoven and they appear either as well-defined broad regions or thin elongated spiral bands. Regions of bounded orbits first and foremost correspond to stability islands of regular orbits where a third adelphic integral of motion is present.
 \item A strong correlation between the extent of the basins of escape/collision and the value of the energy integral was found to exist. Indeed, for low energy levels the structure of both parts of the physical phase space exhibits a large degree of fractalization and therefore the majority of orbits escape randomly. As the value of the energy increases however, several well-formed basins of escape/collison emerge. The extent of these basins of escape is more prominent at high energy levels.
 \item In both parts of the physical space we identified an extremely small portion of trapped chaotic orbits which do not escape within the predefined maximum time of numerical integration. The initial conditions of these orbits are located either in thin chaotic layers (separatrix) inside the regular areas of motion, or near the boundaries of stability islands. Moreover, these orbits reveal their chaotic nature very quickly, so they cannot be considered as sticky orbits.
 \item We observed, that in many cases the escape process is highly sensitive dependent on the initial conditions, which means that a minor change in the initial conditions of an orbit lead the test particle to escape through another direction. These regions are the exact opposite of the escape basins, are completely intertwined with respect to each other (fractal structure) and are mainly located in the vicinity of stability islands. This sensitivity towards slight changes in the initial conditions in the fractal regions implies, that it is impossible to predict through which exit the particle will escape.
 \item Our calculations revealed, that the escape and collision times of orbits are directly linked to the basins of escape and collision, respectively. In particular, inside the basins of escape/collision as well as relatively away from the fractal domains, the shortest escape/collision rates of the orbits had been measured. On the other hand, the longest escape/collision periods correspond to initial conditions of orbits either near the boundaries between the escape/collision basins or in the vicinity of the stability islands. The collision basins wind out as spirals in the outer regions of the plots due to the rotating primaries. However, collision basins at the immediate neighbourhood of the origin have been observed.
  \item Strikingly, the vast majority of the $\dot{\phi} > 0$ part of the physical $(x,y)$ space is covered by initial conditions of orbits that lead to escape from the system. The corresponding escape basins are filed with orbits that leave the system's disk after a short transient time period. Thus, in contrast to the $\dot{\phi} < 0$ part, there appears a border between possible non-escaping orbits and a region where only escape is possible. The total area of stability islands is less for $\dot{\phi} > 0$ than for $\dot{\phi} < 0$.
  \item Collision on the primary body 2 becomes more and more unlikely with decreasing mass ratio. Furthermore, statistically the required time for the orbits in order to escape increases with decreasing mass ratio. In addition, the increasing area of stability islands of bounded motion indicates an increasing tendency toward regularity of the third test body the greater the primary 1, or in other words the smaller mass ratio.
\end{enumerate}

Judging by the detailed outcomes we may say that our task has been successfully completed. We hope that the present numerical analysis and the corresponding results to be useful in the field of escape dynamics in the PCRTBP Hamiltonian. The outcomes as well as the conclusions of the present research are considered, as an initial effort and also as a promising step in the task of understanding the escape mechanism of orbits in this interesting version of the classical three-body problem. Taking into account that our results are encouraging, it is in our future plans to properly modify our dynamical model in order to expand our investigation into three dimensions and explore the entire six-dimensional phase space thus revealing the influence of the value of the energy and of the mass ratio on the orbital structure. Moreover, it would be interesting to apply our numerical methods in some interesting cases of the PCRTBP such the Earth-Moon and the Saturn-Titan systems.

\section*{Acknowledgments}

I would like to express my warmest thanks to J. Nagler and Ch. Jung for all the illuminating and inspiring discussions during this research. My thanks also go to the anonymous referees for the careful reading of the manuscript and for all the apt suggestions and comments which allowed us to improve both the quality and the clarity of the paper.

\section*{Compliance with Ethical Standards}

\begin{itemize}
  \item Funding: The author states that he has not received any research grants.
  \item Conflict of interest: The author declares that he has no conflict of interest.
\end{itemize}

\end{document}